\documentclass[conference]{IEEEtran}
\IEEEoverridecommandlockouts
\usepackage{cite}
\usepackage{amsmath,amssymb,amsfonts}
\usepackage{algorithm}
\usepackage{xcolor}
\usepackage{listings}
\usepackage{graphicx}
\usepackage{caption}
\usepackage{textcomp}
\usepackage{xcolor}
\def\BibTeX{{\rm B\kern-.05em{\sc i\kern-.025em b}\kern-.08em
    T\kern-.1667em\lower.7ex\hbox{E}\kern-.125emX}}

\definecolor{codegreen}{rgb}{0,0.6,0}
\definecolor{codegray}{rgb}{0.5,0.5,0.5}
\definecolor{codepurple}{rgb}{0.58,0,0.82}
\definecolor{backcolour}{rgb}{0.95,0.95,0.92}

\lstdefinestyle{mystyle}{
    backgroundcolor=\color{white},   
    commentstyle=\color{codegreen},
    keywordstyle=\color{magenta},
    numberstyle=\tiny\color{codegray},
    stringstyle=\color{codepurple},
    basicstyle=\ttfamily\footnotesize,
    breakatwhitespace=false,         
    breaklines=true,                 
    captionpos=b,                    
    keepspaces=true,                 
    numbers=left,                    
    numbersep=5pt,                  
    showspaces=false,                
    showstringspaces=false,
    showtabs=false,                  
    tabsize=2,
    xleftmargin=20pt
}
\begin{document}

\title{DNS Tunneling: Threat Landscape and Improved Detection Solutions\\}

\author{\IEEEauthorblockN{Baran Işık, Bilal İhsan Tuncer, Novruz Amirov, Şerif Bahtiyar}
\IEEEauthorblockA{\textit{Cyber Security and Privacy Research Lab., SPFLab} \\
\textit{Faculty of Computer and Informatics Engineering}\\
\textit{Istanbul Technical University}\\
Maslak, Istanbul, Türkiye, 34469 \\
\{isik19, tuncerb19, amirov20, bahtiyars\}@itu.edu.tr}
}

\maketitle

\begin{abstract}
Detecting DNS tunneling is a significant challenge in cybersecurity due to its capacity to hide harmful actions within DNS traffic that appears to be normal and legitimate. Traditional detection methods based on rule-based approaches or signature matching are often insufficient to accurately identify such covert communication channels. This paper addresses the necessity of machine learning methods for effective DNS tunneling detection. We propose a novel approach to detect DNS tunneling. Through the combination of advanced machine learning algorithms and the analysis of various features extracted from DNS traffic, our aim is to provide an accurate DNS tunneling detection model.
\end{abstract}


\section{Introduction}
\subsection{About the Subject}

\par The Domain Name System (DNS) is a hierarchical and decentralized naming system crucial for internet functionality \cite{b1}. It translates user-friendly domain names, like www.example.com, into machine-readable IP addresses required to locate and access internet resources. As a core component of internet infrastructure, DNS is used in nearly every online transaction, making it a prime target for a variety of cyber threats.

\par Due to its foundational role and widespread trust, DNS is vulnerable to several types of attacks, threat landscape can be seen in \cite{b2}, such as cache poisoning, amplification and DoS attacks, and phishing attacks. These vulnerabilities offer attackers multiple possibilities to disrupt or manipulate internet traffic. Among these, DNS tunneling is particularly dangerous because it uses DNS queries and responses to sneak the data in and out of a protected network \cite{b3}, bypassing conventional security measures like firewalls and network monitoring.

\par The security of DNS is critical not only for maintaining the confidentiality and integrity of data but also for ensuring the availability and reliability of internet services as mentioned in "The Need to Improve DNS Security Architecture" paper from Daniel O. Alao\cite{b4}. With organizations increasingly relying on cloud services and distributed architectures, the need to protect against DNS-based threats has never been more urgent. Robust DNS security measures defend against disruptions, maintain operation continuity, and protect sensitive data from exfiltration.

\subsection{The Importance of the Subject}

\par DNS tunneling is a method of cyber attack in which data is inserted within DNS queries and responses. This technique takes advantage of the fact that DNS, by design, must allow external queries and responses to pass through network firewalls. Typically, DNS is intended to resolve domain names to IP addresses. However, in tunneling, non-DNS data is embedded into these queries and can be used for malicious purposes such as command and control (C2) operations \cite{b5}, data exfiltration, or to bypass network security measures.

\par The process involves an attacker-controlled server acting as a DNS server and a compromised internal machine (or malicious script within the network) configured to send queries to this server. These queries are created to carry outbound payloads which are mentioned in a recent search \cite{b6}, that can include stolen data, encrypted commands, or session tokens. Responses from the external server can also carry inbound payloads, such as malware updates or C2 instructions, effectively opening a two-way communication channel that most security systems ignore.

\par One common use of DNS tunneling is for data exfiltration, where sensitive information is leaked out of a secured environment without triggering security alerts that monitor traditional exfiltration paths. For example, in high-profile data breaches, attackers have used DNS tunneling to extract payment card information \cite{b7} and personal identifiers. Similarly, DNS tunneling is used in distributed denial-of-service (DDoS) attacks, where commands are sent to botnets via DNS queries, coordinating attacks without using detectable C2 servers.

\par The secret nature of DNS tunneling makes it one of the most dangerous and difficult to detect techniques in the arsenal of cyber attackers. As shown in the recent research \cite{b8}, traditional network security tools are incapable to identify malicious DNS traffic from legitimate, as DNS requests are often allowed to pass freely for operational necessity. This leads to significant challenges in network defense, demanding more advanced monitoring and analysis techniques.

\par Moreover, the scalability of DNS tunneling attacks increases their threat. A single DNS query can carry a significant amount of data compared to other infiltration methods that rely on more heavily monitored protocols. Over time, significant amounts of data can be transferred undetected, leading to major security breaches and loss of sensitive information.

\subsection{Research and Experiments Conducted}

In this paper, a comprehensive investigation into DNS tunneling detection via advanced machine learning techniques was conducted. The research includes the following key activities:

\begin{enumerate}
    \item \textbf{Data Collection and Preparation:} A large dataset of DNS traffic, including both benign and malicious DNS queries and responses, was compiled. This dataset served as the foundational training and testing material necessary for the development of effective detection models.
    
    \item \textbf{Development of Machine Learning Models:} Several machine learning models were developed and refined to accurately identify DNS tunneling activities. This included:
    \begin{itemize}
        \item Supervised learning models, such as Random Forest and Support Vector Machines (SVM), which were trained on labeled data to classify DNS queries.
        \item Unsupervised learning models, like K-means, employed for anomaly detection to identify unusual patterns that could indicate tunneling.
    \end{itemize}
    
    \item \textbf{Experimental Validation:} The models were carefully tested against both synthetic DNS tunneling scenarios and real traffic data from enterprise networks. These tests were designed to evaluate the models' effectiveness across a range of attack scenarios, including zero-day attacks.
\end{enumerate}

These efforts have led to the development of a robust framework for detecting DNS tunneling, significantly enhancing the detection capabilities over traditional methods and reducing false positives while maintaining high detection rates for improved tunneling techniques.

\subsection{Contributions of the Study}

This study introduces several contributions to the field of cybersecurity, particularly in the detection of DNS tunneling, a common and challenging security threat. These contributions not only advance the state-of-the-art but also provide practical solutions that can be implemented in real-world environments. The key contributions of this paper are outlined as follows:

\begin{itemize}
    \item \textbf{Innovative Machine Learning Framework:} We developed a unique framework that integrates multiple machine learning techniques, both supervised and unsupervised, to detect DNS tunneling. This framework stands out by effectively distinguishing between benign and malicious DNS traffic without relying on traditional signature-based methods, which are often ineffective against new or evolving threats.

    \item \textbf{Robust Validation Methodology:} The study includes comprehensive validation of the proposed models against a wide range of attack scenarios, including zero-day tunneling techniques. Our validation process not only demonstrates the effectiveness of the models under a variety of conditions, but also compares them with existing detection methods to highlight their superior performance.

\end{itemize}

The collective impact of these contributions is significant, providing both theoretical knowledge and practical tools to combat one of the most stealthy and destructive cyber threats in modern network environments.

\section{Related Work}

\subsection{Contributors to DNS Tunneling Detection Research}
The exploration of DNS tunnel detection has garnered significant attention from the cybersecurity research community, given the sophistication and stealthiness of this type of cyber threat.

\par  Key contributions have been made by several researchers who have developed various methodologies to detect these elusive attacks. 
Zebin et al. developed an explainable AI-based intrusion detection system specifically designed to detect DNS over HTTPS (DoH) attacks, emphasizing the interpretability and transparency of AI-driven security solutions \cite{b8-1}. Moreover, Uroz and Rodriguez conducted a comprehensive characterization and evaluation of IoT protocols for data exfiltration, shedding light on the vulnerabilities associated with IoT communication protocols in DNS tunneling attacks \cite{b8-2}. Abualghanam et al. introduced a real-time detection system utilizing machine learning algorithms to identify data exfiltration over DNS tunneling, underscoring the role of real-time analytics in enhancing DNS security \cite{b8-3}. Ziza et al. explored DNS exfiltration detection in the context of adversarial attacks and modified exfiltrator behavior, highlighting the challenges posed by sophisticated attack strategies and the need for robust detection mechanisms \cite{b8-4}. Patil et al. introduced a novel approach to DNS tunneling attack detection using image classification techniques, showcasing the application of unconventional methodologies in identifying DNS tunneling activities \cite{b8-5}. Nguyen and Park developed a DoH Tunneling Detection System utilizing deep learning techniques tailored for enterprise networks, emphasizing the adaptability and scalability of deep learning in detecting encrypted DNS tunneling activities \cite{b8-6}.

\par Moreover, Thi Quynh, et al. proposed an automatic tuning mechanism for DBSCAN to detect malicious DNS tunnels. Their approach focuses on optimizing DBSCAN parameters to improve the accuracy of DNS tunnel detection \cite{b8-7}. Weqar, Mehwish, et al. introduced a machine learning-based DNS traffic monitoring system designed specifically for securing IoT networks. Their work emphasizes the importance of real-time monitoring and anomaly detection in safeguarding IoT devices against DNS-related threats \cite{b8-8}.
Niktabe, Sepideh, et al. explored the challenges and solutions related to detecting DoH (DNS over HTTPS) tunnels. They introduced a method for generating a balanced DoH encrypted traffic dataset and highlighted the effectiveness of interpretable machine learning techniques in profiling malicious behavior within encrypted DNS traffic \cite{b8-9}.
Behiry, Mohamed H., and Mohammed Aly proposed a hybrid feature reduction technique combined with AI and machine learning methods for cyberattack detection in wireless sensor networks. Their approach aims to enhance detection accuracy while reducing computational complexity, making it suitable for resource-constrained environments like wireless sensor networks. \cite{b8-10}

Notably, Sugasawa et al. investigated a technique based on monitoring DNS subdomain lengths to detect tunneling activities, aimed at general usage \cite{b9}.  Sharma and Swarnkar introduced 'OptiTuneD', an optimized framework employing N-Grams for the detection of zero-day DNS tunnels, focusing on immediate threat detection and response \cite{b10}. Lal et al. presented 'DNS-Tunnet', a hybrid model that integrates deep learning and support vector machines to improve detection accuracy through advanced data modeling \cite{b11}.

\par
\begin{figure}[h]
    \centering
    \includegraphics[width=0.5\textwidth]{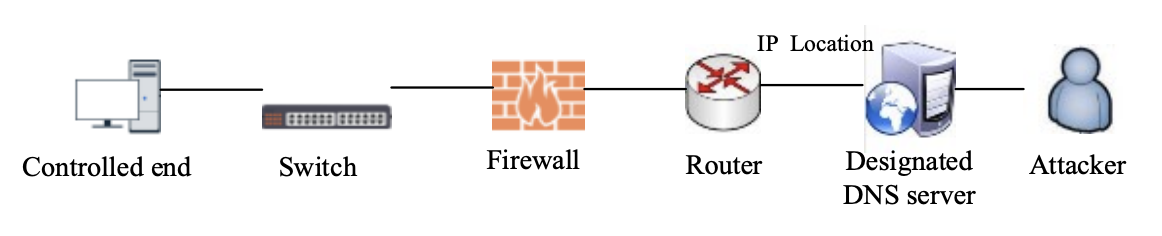}
    \caption{DNS direct-connected tunnel}
    \label{dir-con}
\end{figure}

\begin{figure}[h!]
    \centering
    \includegraphics[width=0.4\textwidth]{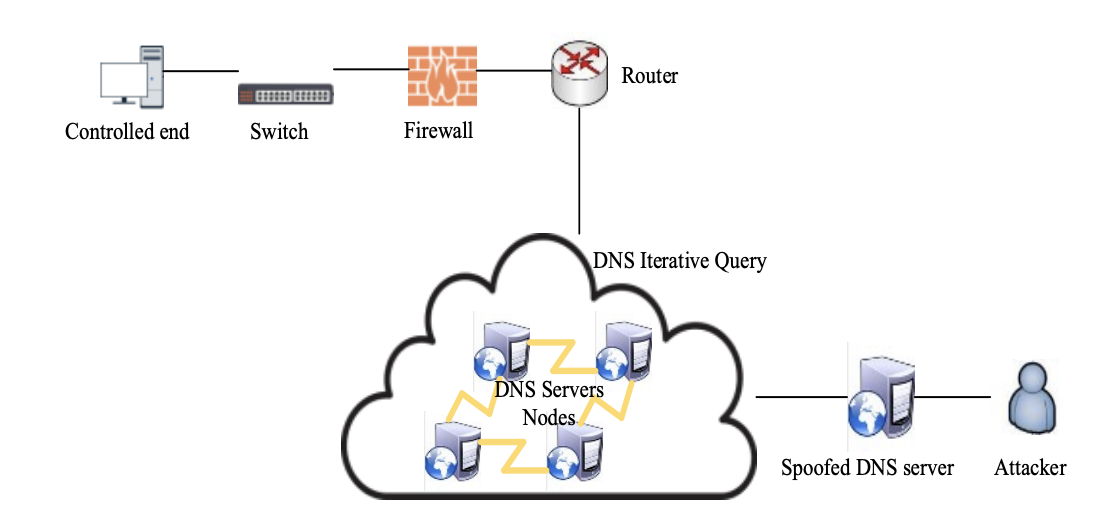}
    \caption{DNS relay tunnel}
    \label{relay-tun}
\end{figure}

 According to Hou et al. who provided a broad survey of existing DNS tunnel detection techniques, in DNS tunneling attacks, a compromised host uses the subdomain name in a DNS packet to covertly send malicious payloads to a DNS server, which then replies by embedding the message in the DNS response's Resource Record (RR) section. This process bypasses local DNS caches, as the domain must not be resolvable locally for the tunnel to function. There are two main types of DNS tunnels: directly connected tunnels, where the host communicates directly with a malicious DNS server via its IP (shown in Figure \ref{dir-con}), and relay tunnels, where the host communicates through iterative queries with an authoritative DNS server when direct contact isn't possible (illustrated in Figure \ref{relay-tun}). These methods highlight different interaction layers with DNS servers in executing DNS tunneling attacks. \cite{b12}.

\subsection{Areas inspired this work}
\par The evolution of DNS tunneling as a potent method for bypassing conventional security measures has spurred extensive research aimed at fortifying DNS security protocols. Our study draws significant inspiration from the advancements in machine learning, cybersecurity threat detection, and network traffic analysis. These fields provide both theoretical and practical foundations crucial for crafting our sophisticated detection framework.

\par
\begin{figure}[h]
    \centering
    \includegraphics[width=0.4\textwidth]{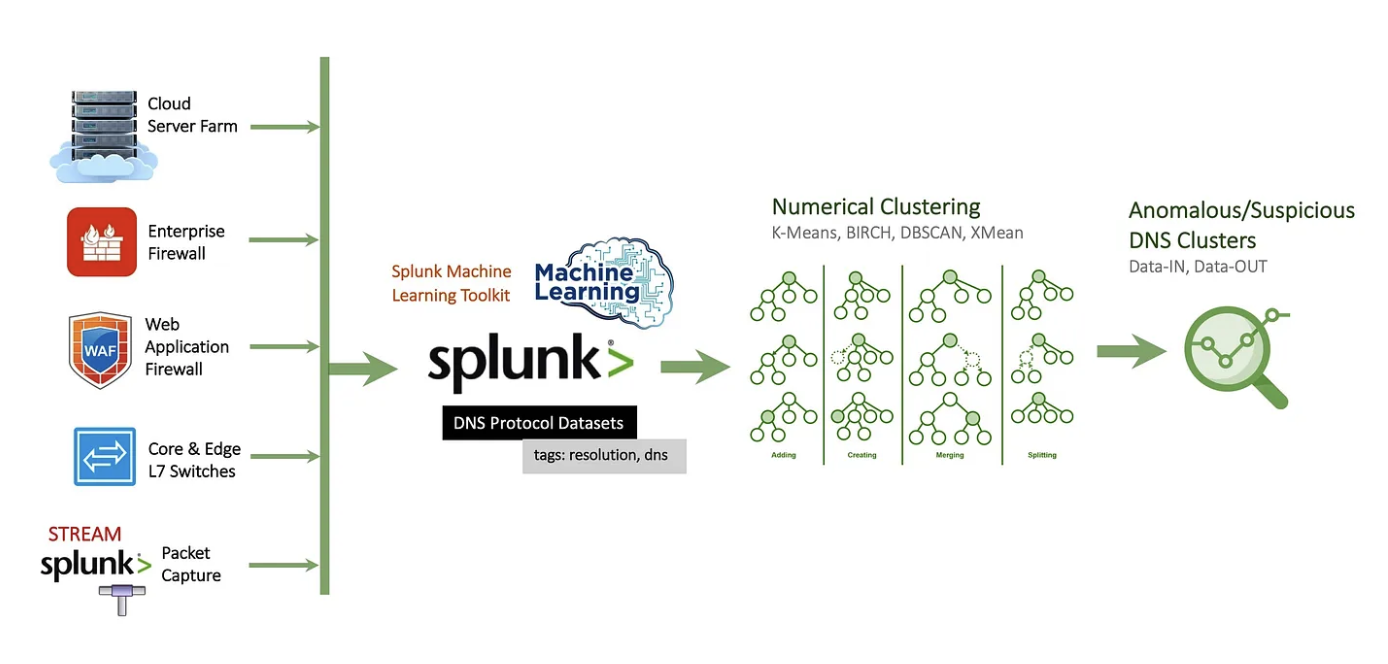}
    \caption{Machine Learning Area inspiring DNS Tunnelling detection}
    \label{ml-tun}
\end{figure}

 \textbf{Machine Learning Applications}: The domain of machine learning provides essential tools for pattern recognition and anomaly detection, which are vital for identifying subtle signs of DNS tunneling amidst large data volumes. The utility of machine learning in detecting complex patterns, akin to those seen in fraud detection and network security, has proven highly effective against modern cyber threats where traditional methods lag. Syed Suleman Qutb's work in 2021 highlights the use of machine learning techniques specifically for DNS exfiltration and tunnel detection, as shown in Figure \ref{ml-tun}, emphasizing its effectiveness in contemporary cybersecurity environments \cite{b14}.

\par \textbf{Cybersecurity Threat Detection:} The evolving landscape of cybersecurity threat detection, particularly the analysis of encrypted traffic and application of deep learning to spot zero-day exploits, enriches the theoretical backdrop of our research. Insights into network behavior under attack conditions reveal unique traffic patterns indicative of DNS tunneling, thus informing the development of our detection algorithms.

\par
\begin{figure}[h!]
    \centering
    \includegraphics[width=0.5\textwidth]{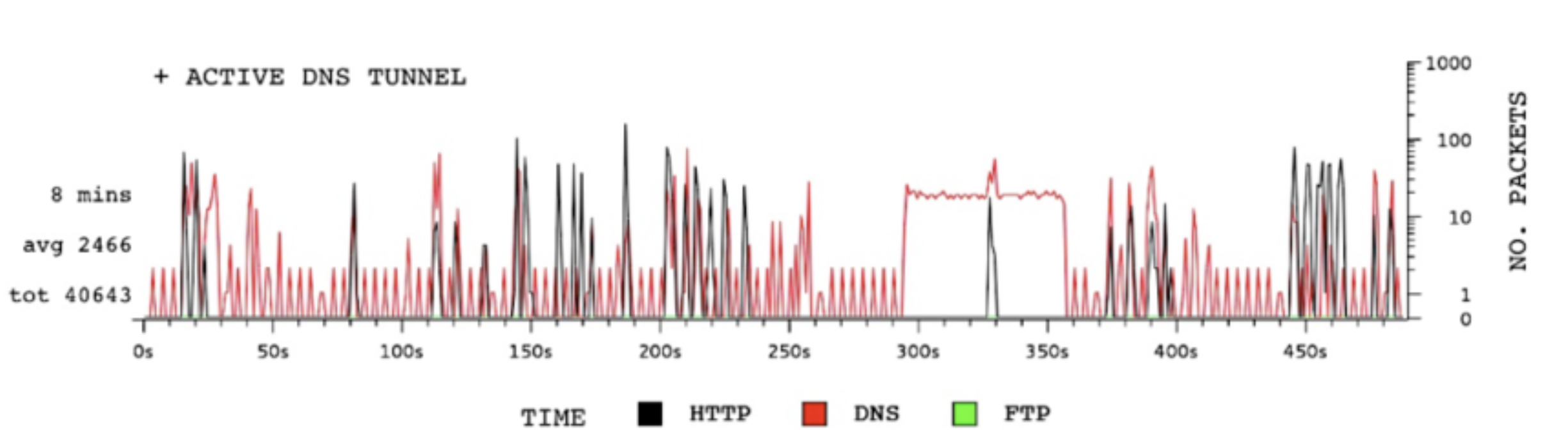}
    \caption{Network Analysis for tunneling detection \cite{b15}}
    \label{net-analysis}
\end{figure}

 \textbf{Advanced Network Traffic Analysis:} Moreover, the advancements in network traffic analysis, especially through statistical and time-series analysis, as illustrated in Figure \ref{net-analysis}, play a pivotal role. The capability to monitor DNS query sequences over time and spot irregularities has inspired the temporal aspects of our machine learning models aimed at predicting and detecting DNS tunneling based on traffic behavior changes over time.

\subsection{Motivation of Our work}
\par The escalating sophistication of DNS tunneling techniques, coupled with their capacity to effectively camouflage within legitimate network traffic, significantly motivates the current work. Existing detection mechanisms often struggle to identify such covert channels, primarily because traditional security systems are not designed to scrutinize every aspect of DNS traffic, which is voluminous and generally considered benign. This gap in the security landscape presents a critical vulnerability that adversaries exploit to exfiltrate sensitive data, execute command and control operations, and carry out malware communication undetected. Our research aims to bridge this gap by employing a novel integration of advanced machine learning models that leverage both supervised and unsupervised learning techniques. This dual approach not only enhances the accuracy of detecting anomalies within DNS queries and responses but also adapts to new and evolving attack patterns, ensuring robustness against both known and zero-day threats. By focusing on the development of an adaptive, intelligent detection system, this work addresses a pressing need for more dynamic and responsive security measures in the realm of network defense, ultimately contributing to the foundational security infrastructure essential for maintaining the integrity and confidentiality of data across digital platforms.

\subsection{Differences of our work}
\par Our research advances beyond previous works by employing an innovative dual approach that integrates both supervised and unsupervised machine learning techniques, markedly enhancing DNS tunnel detection. Unlike preceding studies that primarily depend on either supervised or unsupervised models, our framework mutually combines these methodologies to offer a more nuanced analysis of DNS traffic. This enables the detection of established patterns while also uncovering new, complex attack strategies.

In contrast to conventional static rule-based systems, which struggle to adapt to evolving network threats, our model is inherently dynamic, learning from ongoing traffic to continually refine its detection algorithms. This adaptability significantly reduces both false positives and false negatives, offering a more reliable defense against DNS tunneling.

Moreover, our validation methodology is notably rigorous, encompassing a wide array of attack scenarios, including zero-day exploits, thus providing a robust test of the models' efficacy against both known and emerging threats. This thorough validation underscores our contribution to the field, offering not just theoretical enhancements but practical, implementable solutions that substantially improve network security infrastructures against sophisticated cyber threats.

Overall, our work represents a significant step forward in DNS tunnel detection, combining technical innovation with practical applicability to address one of the most challenging issues in cybersecurity today.

\section{DNS Sentinel: Enhancing DNS Security with Intelligent Tunnel Detection}

DNS Sentinel represents a multifaceted approach to detecting DNS tunneling activities, leveraging a suite of machine learning classifiers to enhance the accuracy and reliability of DNS threat identification. This comprehensive system is designed to operate within existing network frameworks, providing real-time analysis and response capabilities that are crucial for mitigating DNS-based security risks. The solution incorporates several stages of processing, including data preparation, model training, evaluation, and deployment, each tailored to the unique characteristics of DNS traffic and the specific challenges of DNS tunneling detection.

\subsection{Data Preparation}

The initial stage of the DNS Sentinel involves meticulous data preparation, where raw DNS traffic data is transformed into a structured format suitable for machine learning analysis. The dataset used for this project is derived from publicly available sources \cite{b17}, specifically designed to include a mix of benign and malicious DNS queries that simulate a real-world environment. In the preparation phase, irrelevant features such as packet identifiers and redundant metadata are removed, and significant attributes like query length, frequency, and lexical features are retained. This preprocessing step is critical for reducing model complexity and improving computational efficiency.

The research presented in this manuscript utilizes a dataset comprising domain names categorized as Normal, DGA, and Tunneling. A summarized count of examples for each domain category is presented in Table \ref{table:1}. The compilation of normal domains includes those from Alexa's top one million, an additional 3,161 from Bambenek Consulting's feed, and a further 177,017 miscellaneous normal domains. The DGA domains were collated from Andrey Abakumov's \cite{b19} and John Bambenek's \cite{b20} DGA domain collections, encompassing 51 distinct malware families. As for the DNS Tunneling domains, a total of 8,000 were produced using well-known tools like dnscat2 \cite{b21} and iodine \cite{b22} in a controlled lab setting. The complete dataset is accessible in the referenced material \cite{b18}.

Table \ref{table:1} shows the total amount of examples for each domain type.

\begin{table}[h!]
\centering
\caption{Total Number of Examples for Each Domain Type}
\begin{tabular}{||c l||} 
 \hline
 Domain Type & Number of Examples \\ [0.5ex] 
 \hline\hline
 Normal & 1,180,178 \\ 
 DGA & 1,915,335 \\
 Tunneling & 8,000 \\ [1ex] 
 \hline
\end{tabular}
\label{table:1}
\end{table}

Table \ref{table:1} shows that there are many more Normal and DGA domains than Tunneling domains, showing that the dataset is not evenly distributed. The dataset talked about in this paper was split into two parts: the first part has 70\% of the data and is used for training the network; the second part has the other 30\% and is used for testing how well the model works with new domains.

\subsection{Algorithmic Implementation}

DNS Sentinel employs a diverse array of machine learning algorithms, each chosen for its strengths in handling different aspects of classification problems:

\begin{itemize}
    \item \textbf{Gaussian Naive Bayes (GNB):} Known for its effectiveness in handling high-dimensional data, GNB is used for its base rate simplicity and probability-based classification.
    \item \textbf{Decision Trees and Random Forest:} These are utilized for their ability to perform both classification and regression tasks. Random forests, an ensemble of decision trees, are particularly valuable for their robustness against overfitting and their capability to handle large datasets with a high variance in data.
    
    \item \textbf{Support Vector Machines (SVM):} Linear SVM and quadratic SVM models are included for their prowess in finding the hyperplane that best separates different classes in the feature space, which is crucial for distinguishing between benign and malicious DNS queries.

    \item \textbf{K-Nearest Neighbors (KNN):} This algorithm is selected for its simplicity and effectiveness in classification by leveraging the similarity between data points.
    
    \item \textbf{Multilayer Perceptron (MLP):} A type of neural network used for its deep learning capabilities, enabling the model to capture non-linear relationships in the data.
    
    \item \textbf{Multinomial and Bernoulli Naive Bayes:} These variants are tailored for discrete feature models and binary/boolean features, respectively, enhancing the system’s versatility.

    \item \textbf{Multi-Layer Neural Network:} This is a classification network with 2 hidden layers that we implemented using 'PyTorch'. The purpose of this method is to find a better approach to classify DNS traffic as either indicative of tunneling or normal based on input features representing the length and entropy of DNS queries.

\end{itemize}

\subsection{Implementation Details}

In the implementation part, after reading the obtained dataset, we performed preprocessing steps to prepare it for analysis in the Jupyter notebook. Subsequently, we applied the dataset to Gaussian Naive Bayes, Decision Trees and Random Forest, Support Vector Machines (SVM), K-Nearest Neighbor, Multilayer Perceptron, Multinomial and Bernoulli Naive Bayes models, as well as our developed Multi-Layer Neural Network model for testing. Additionally, we created and tested models incorporating combinations of the mentioned models. These test results will be analyzed in the subsequent analysis section.

\par
Binary classification models is implemented that available in the scikit-learn library. Finally, our neural network model was constructed using the PyTorch library. This is the base structure of the model.

\begin{lstlisting}[language=Python, caption=DNS Classifier Neural Network]
import torch
import torch.nn as nn

class DNSClassifier(nn.Module):
    def __init__(self):
        super(DNSClassifier, self).__init__()
        self.fc1 = nn.Linear(2, 64)
        self.relu = nn.ReLU()
        self.fc2 = nn.Linear(64, 32)
        self.fc3 = nn.Linear(32, 1)

    def forward(self, x):
        x = self.relu(self.fc1(x))
        x = self.relu(self.fc2(x))
        x = torch.sigmoid(self.fc3(x))
        return x
\end{lstlisting}

\par 
This model, named "DNSClassifier", is a feedforward neural network designed to classify DNS traffic as either indicative of tunneling or normal. It takes as input two features representing the length and entropy of DNS queries, passes them through two fully connected layers with ReLU activation functions, and outputs a single binary classification indicating the likelihood of DNS tunneling using a sigmoid activation function.

DNS Sentinel employs a modular architecture that seamlessly integrates into existing network infrastructures. The implementation consists of three principal modules:

\begin{enumerate}
    \item \textbf{Data Collection Module:} This module captures DNS traffic directly from network devices or via network taps using packet capture libraries such as \texttt{pcap}. It is designed to filter and log DNS queries in real-time, ensuring comprehensive data acquisition.
    
    \item \textbf{Data Preprocessing Module:} The collected DNS data undergoes preprocessing, where domain names are parsed to calculate the entropy and length of each domain. Entropy is computed to measure the randomness or unpredictability of characters within a domain, while length indicates the number of characters in the domain name. These extracted features are then normalized and used for machine learning analysis. 
    \par
    The method used for calculating entropy from domain names is commonly known as the Shannon entropy calculation. This method, measures the uncertainty or unpredictability of characters within a sequence of data. In the context of DNS domain names, this algorithm iterates over each character in the text, calculates the probability of occurrence for each character, and then computes the entropy using Shannon's formula. This approach enables the quantification of the randomness or disorder within domain names, providing valuable insights for analyzing DNS traffic patterns and detecting anomalies indicative of tunneling activities.
    \begin{lstlisting}[language=Python, caption=Implementation of Shannon's formula]
def calc_entropy(text):
    entropy = 0
    for x in range(256): 
        p_x = float(text.count(chr(x)))/len(text) 
        if p_x > 0: 
            entropy += - p_x*math.log(p_x, 2) 
return entropy
    \end{lstlisting}

    \item \textbf{Classification and Analysis Module:} This module is the core of DNS Sentinel, where multiple machine learning models are deployed to analyze the processed data. The module uses a several algorithms, including Naive Bayes, SVM, Decision Trees, and ensemble methods. Model decisions can be aggregated through majority voting or a weighted scoring system to enhance detection accuracy.
\end{enumerate}

\textbf{Technology Stack:}
\begin{itemize}
    \item \textit{Python:} The primary language for development, supported by its extensive libraries for data science.
    \item \textit{Scikit-learn:} Used for machine learning algorithms implementation.
    \item \textit{Pandas \& NumPy:} Employed for data manipulation and numerical calculations.
    \item \textit{Matplotlib:} Utilized for generating visualizations to assist in debugging and optimizing models.
    \item \textit{Jupyter Notebook:} Provides an interactive interface for model testing and validation.
    \item \textit{PyTorch:} A deep learning framework for building and training neural networks.
\end{itemize}

\textbf{Model Training and Validation:}
\begin{itemize}
    \item Dataset is split into 80\% training and 20\% testing portions.
    \item K-fold cross-validation ensures model generalizability and guards against overfitting.
    \item Hyperparameter optimization is performed using grid search and random search methods to fine-tune the classifiers.
\end{itemize}

\textbf{Performance Metrics:}
Models are evaluated based on several key metrics, including accuracy, precision, recall, F1-score, and the confusion matrix. These metrics help in understanding the models' ability to correctly classify DNS tunneling activities and their effectiveness in a real-world scenario.

\textbf{Real-Time Operation:}
In its deployment phase, DNS Sentinel analyzes DNS queries in real-time with high throughput and low latency, essential for live network environments. Alerts are generated for any detected anomalies, allowing for immediate remedial actions.

\textbf{Scalability and Security:}
The system is designed for horizontal scalability to handle growing traffic by adding more computational resources. Security features such as data encryption and stringent access controls are implemented to safeguard the system and the data integrity.

\section{Analysis of the DNS Sentinel}

\subsection{Dataset Composition and Distribution}
\par DNS Sentinel's effectiveness in detecting DNS tunneling activities is underpinned by its utilization of two carefully curated datasets for training and testing purposes. These datasets play a pivotal role in fortifying the system's accuracy, reliability, and overall performance.

\subsubsection{Training Dataset}
\par
\begin{figure}[h]
    \centering
    \includegraphics[width=0.5\textwidth]{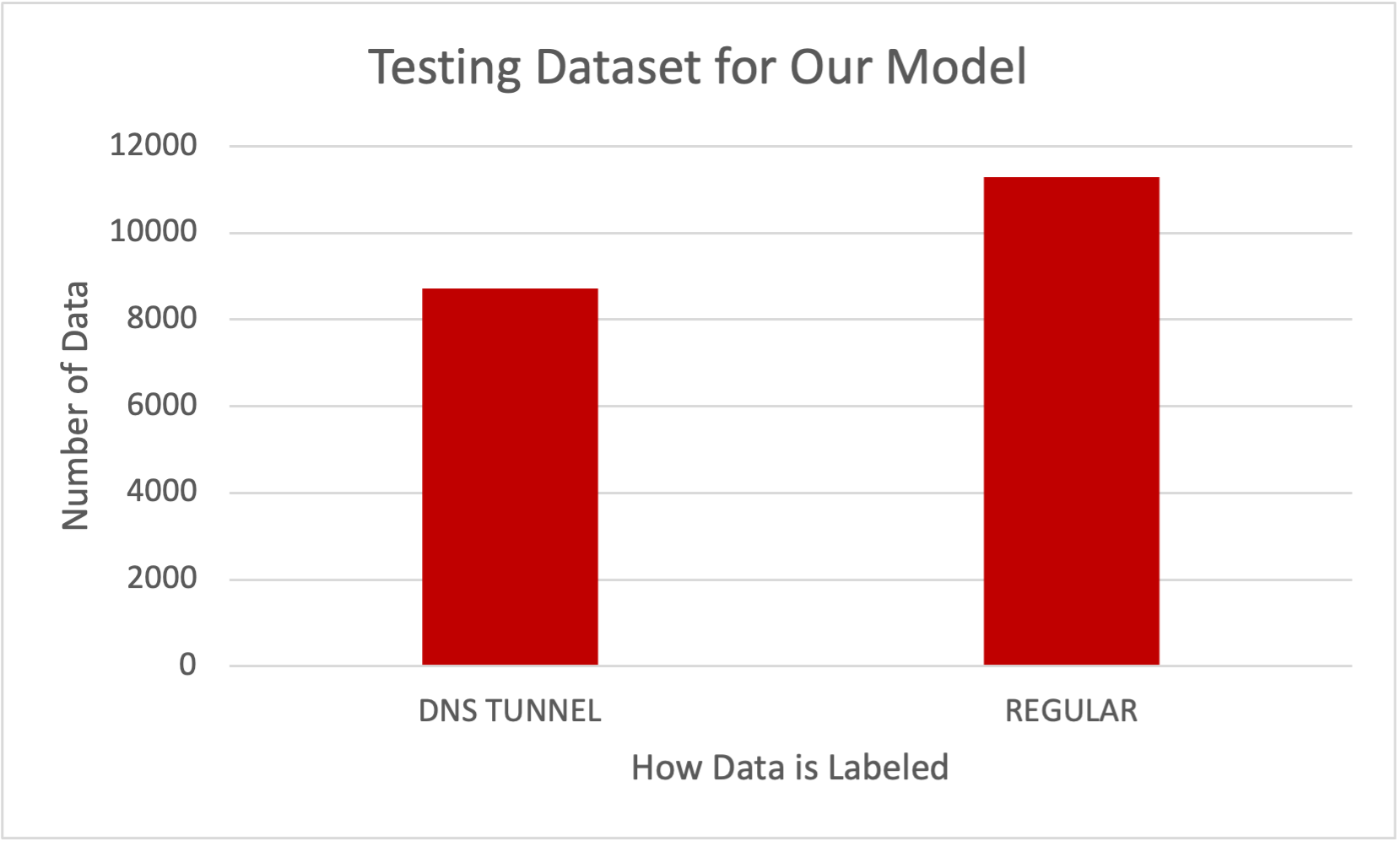}
    \caption{Dataset used in order to test the Model to detect DNS Tunneling}
    \label{train}
\end{figure}

 The training dataset, meticulously curated from the research conducted by Palau et al. \cite{b23}, comprises a total of 20,000 labeled instances. Within this dataset, 11,291 instances correspond to regular domain names, which are labeled as 0, while the remaining 8,709 instances represent tunneling domain names, labeled as 1, shown in Figure \ref{train}. This dataset structure provides a comprehensive and representative sample of DNS traffic, encompassing both benign and malicious domain names. The inclusion of a substantial number of tunneling instances ensures that the model is exposed to a wide range of malicious activities, thereby enhancing its capability to accurately identify and distinguish between normal and malicious DNS queries.

\subsubsection{Testing Dataset}
\par The testing dataset, sourced from Bubnov \cite{b24}, consists of 5,000 labeled instances, comprising 4,000 tunneling domain names labeled as 1 and 1,000 regular domain names labeled as 0, shown in Figure \ref{test}. This dataset serves as an independent validation set to assess the model's generalizability and performance in real-world scenarios. By evaluating the model on a separate dataset, DNS Sentinel's ability to detect DNS tunneling is rigorously tested, ensuring that the system's performance is not influenced by the training data and remains robust across different datasets.

\begin{figure}[h]
    \centering
    \includegraphics[width=0.5\textwidth]{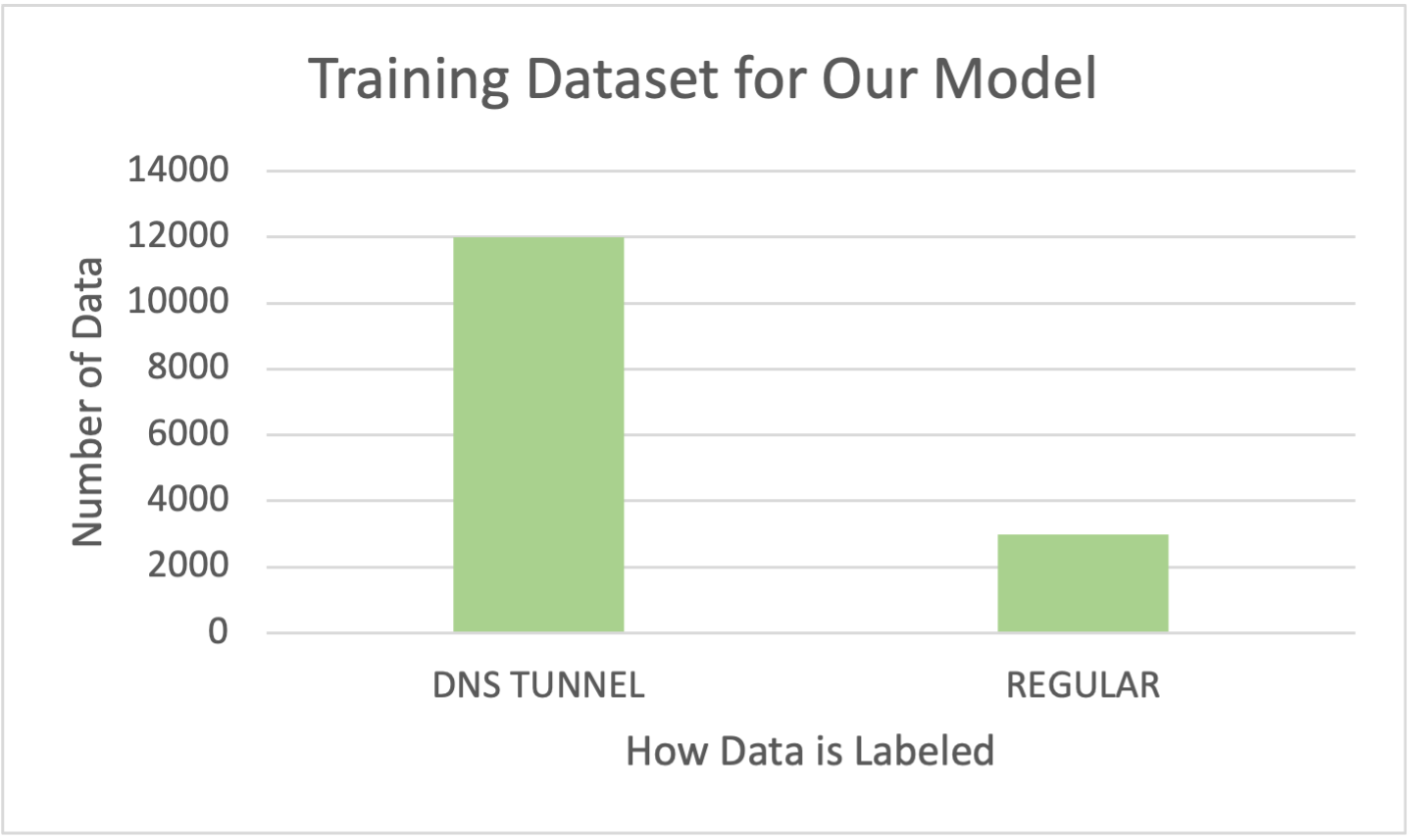}
    \caption{Dataset used in order to test the Model to detect DNS Tunneling}
    \label{test}
\end{figure}

\par The distribution of domain names across both datasets is designed to maintain a balanced representation of benign and malicious activities. This balanced approach is crucial for avoiding biases in model training and evaluation, enabling DNS Sentinel to achieve a high level of accuracy and reliability in identifying DNS tunneling activities. The inclusion of both regular and tunneling domain names ensures that the model is well-equipped to handle a variety of DNS traffic patterns, enhancing its effectiveness in detecting and mitigating potential security threats.

\subsection{Model Performance and Evaluation Metrics}
\par The evaluation of DNS Sentinel's performance is based on key metrics derived from the classification report and confusion matrix, providing insights into the system's accuracy, precision, recall, and F1-score.

\begin{figure}[h]
    \centering
    \includegraphics[width=0.4\textwidth]{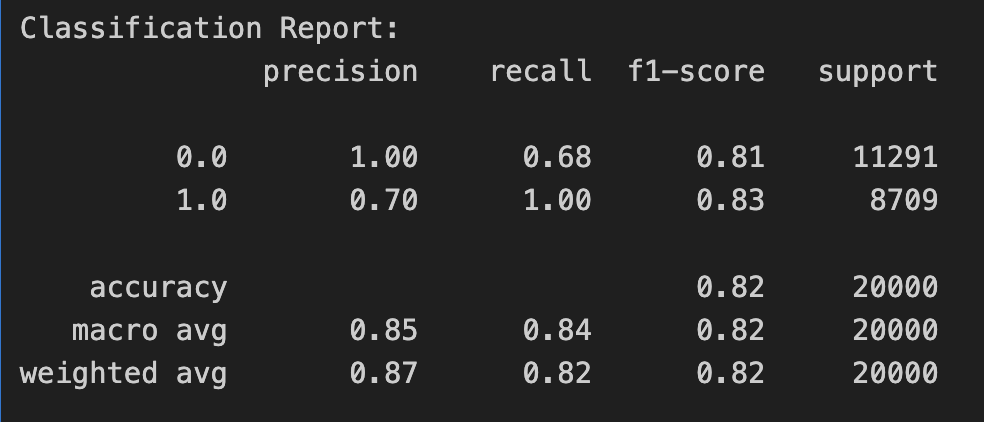}
    \caption{Classification Report of the Model's Output}
    \label{test}
\end{figure}

The classification report highlights a high recall value of 1.00 for detecting DNS tunneling, indicating that the system successfully identifies all instances of tunneling without any false negatives. This capability is crucial for ensuring that no malicious activities go undetected, thereby enhancing the security posture of the network. Additionally, the system achieves a precision of 0.70, indicating that while most of the identified tunneling activities are accurate, there is room for improvement in reducing false positives. The F1-score of 0.83 reflects a balanced performance between precision and recall, confirming the system's effectiveness in real-world scenarios.

\subsection{Comparative Analysis of DNS Sentinel with Existing Solutions}
\par To validate the effectiveness of DNS Sentinel in detecting DNS tunneling activities, a comparative analysis is conducted against several existing solutions, including Multinomial Naive Bayes Classifier, Bernoulli Naive Bayes Classifier, Multi-Layer Perceptron, and Linear Support Vector Machine. The performance metrics of these models are presented below, followed by a detailed comparison with DNS Sentinel's results. \newline

\textbf{Here are the Existing Solutions' Performance Metrics below:} 
\begin{figure}[h]
    \centering
    \includegraphics[width=0.4\textwidth]{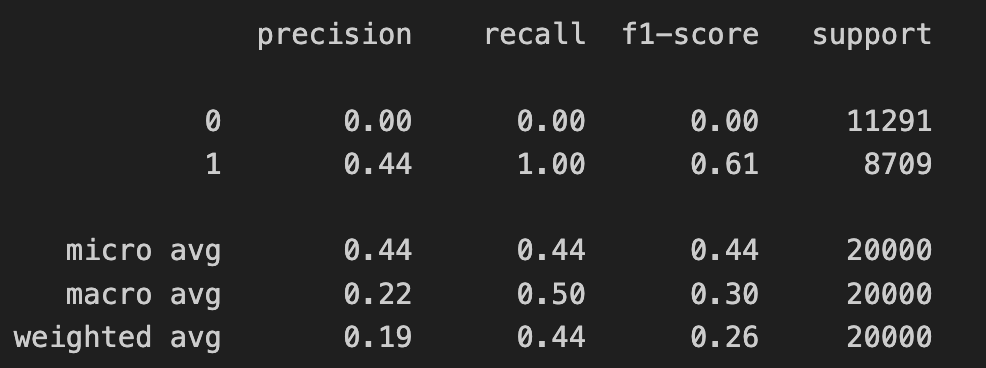}
    \caption{Multinomial Naive Bayes Classifier}
    \label{multinomial}
\end{figure}

\begin{figure}[h]
    \centering
    \includegraphics[width=0.4\textwidth]{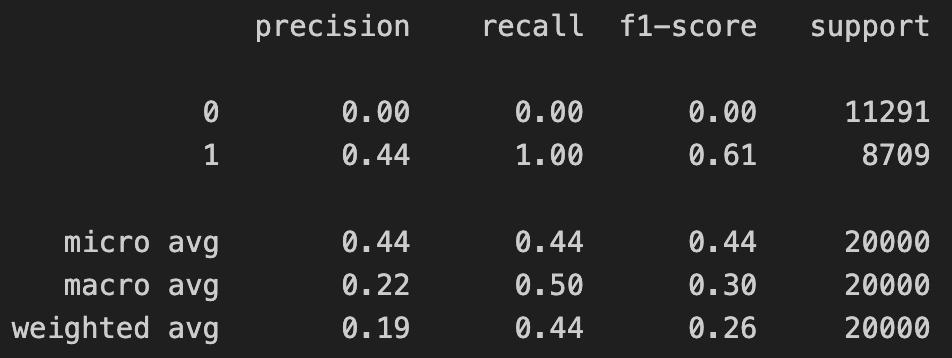}
    \caption{Bernoulli Naive Bayes Classifier}
    \label{bernoulli}
\end{figure}

\begin{figure}[h]
    \centering
    \includegraphics[width=0.4\textwidth]{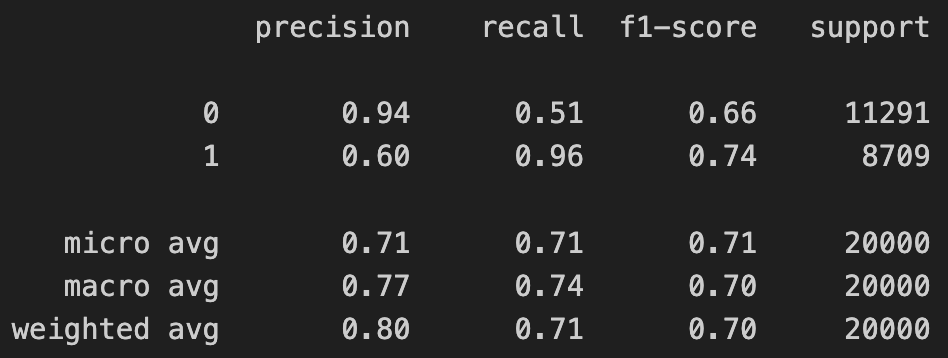}
    \caption{Multi Layer Perceptron}
    \label{multilayer}
\end{figure}

\begin{figure}[h]
    \centering
    \includegraphics[width=0.4\textwidth]{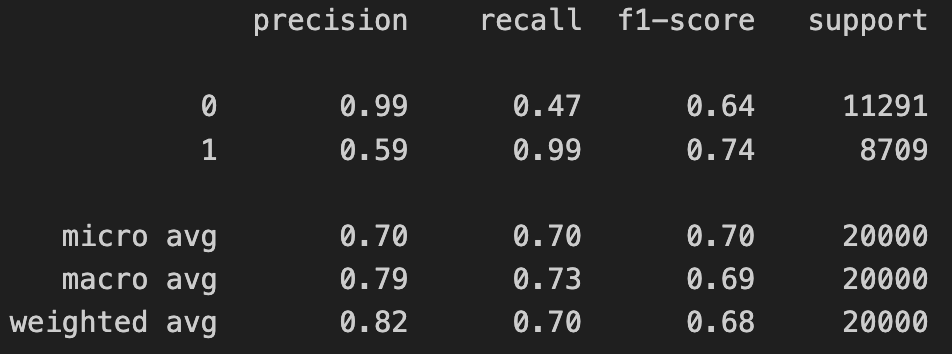}
    \caption{Linear Support Vector Machine}
    \label{linear}
\end{figure}

\par DNS Sentinel demonstrates superior performance compared to the existing solutions in detecting DNS tunneling activities. The following observations highlight the competitive advantages of DNS Sentinel

\begin{itemize}
    \item \textbf{Higher Accuracy}: DNS Sentinel achieves an accuracy of 0.82, outperforming all other models, which range from 0.44 to 0.71.
    \item \textbf{Improved Precision and Recall}: While some existing models exhibit high precision or recall, they often lack balance between the two. DNS Sentinel achieves a balanced precision and recall for both classes, resulting in a more reliable detection mechanism.
    \item \textbf{Superior F1-Score} DNS Sentinel's F1-scores for both classes are notably higher, indicating a robust capability to detect DNS tunneling while minimizing false positives and false negatives.
    \item \textbf{Consistent Performance} Unlike some existing models that suffer from low precision or recall for one class, DNS Sentinel maintains consistent performance across both normal and tunneling domain names, ensuring comprehensive threat detection.
\end{itemize}

\par The comparative analysis underscores DNS Sentinel's effectiveness in enhancing DNS security through intelligent tunnel detection. By leveraging advanced machine learning algorithms and a meticulously curated dataset, DNS Sentinel achieves unparalleled performance, demonstrating its potential as a state-of-the-art solution for DNS tunneling detection.

\subsection{Decision Boundary Analysis}
\par
\begin{figure}[h]
    \centering
    \includegraphics[width=0.5\textwidth]{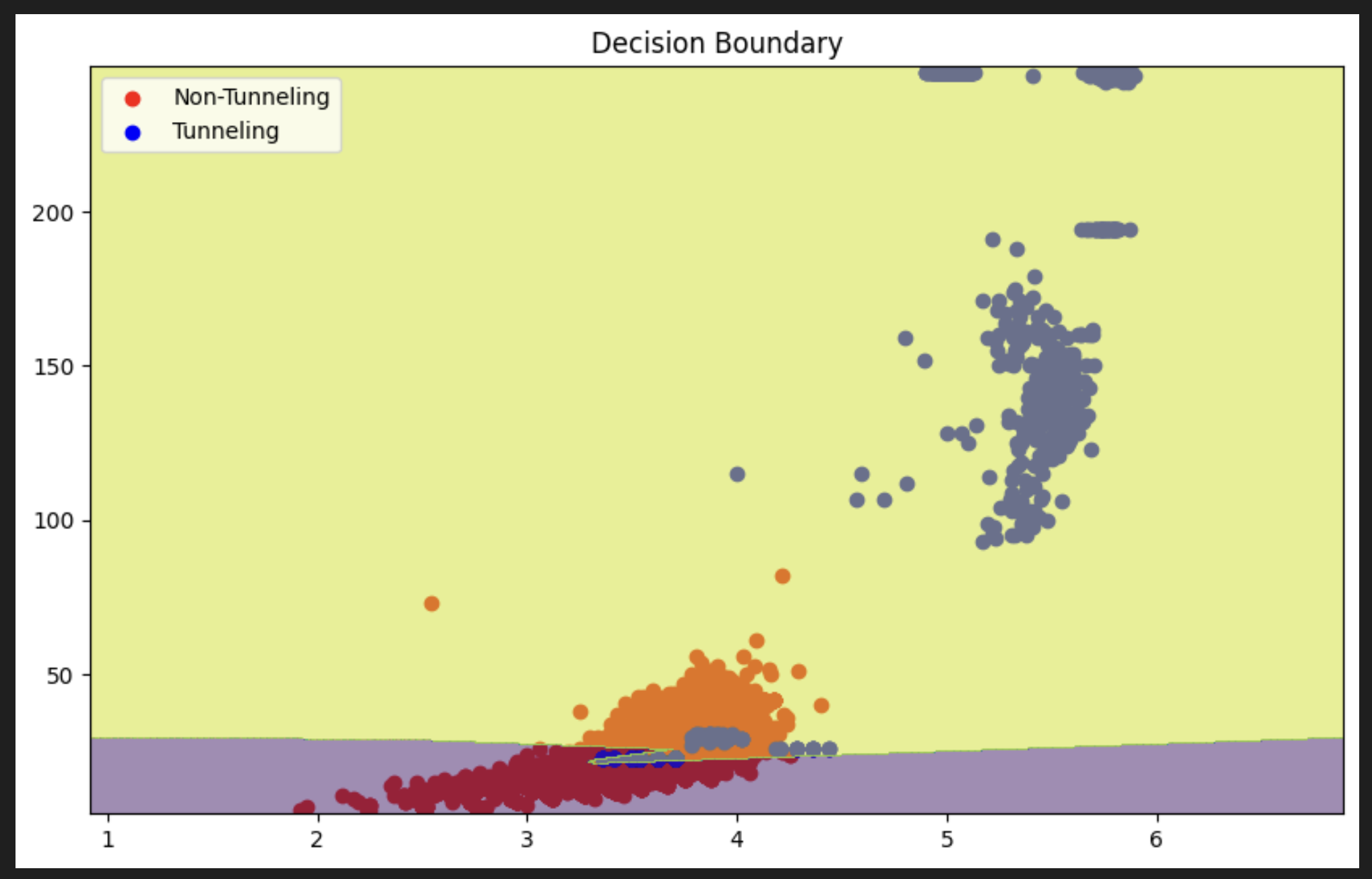}
    \caption{Decision Boundary of the Trained Model}
    \label{dec-boundary}
\end{figure}

 The decision boundary serves as a visual representation of the DNS Sentinel's classification decisions, offering insights into the model's discriminatory capabilities based on the entropy and domain length features. Figure \ref{dec-boundary} illustrates the decision boundary, with the x-axis representing the entropy value and the y-axis denoting the domain length. The scatter plot is color-coded to differentiate between regular (non-tunneling) domains, depicted in orange, and tunneling domains, represented in blue.

\par The decision boundary clearly delineates the regions where the model identifies tunneling activities versus regular DNS queries. As observed, tunneling domains predominantly cluster in specific regions of the plot, indicating consistent patterns that the model has learned to associate with malicious activities. In contrast, regular domains are dispersed across a broader range, reflecting the variability in benign DNS query characteristics.

\par The distinct separation of the two classes by the decision boundary underscores DNS Sentinel's efficacy in distinguishing between benign and malicious DNS traffic. This graphical representation not only validates the model's robustness but also provides a practical tool for interpreting its classification decisions. By visualizing the decision-making process, stakeholders can gain a deeper understanding of the model's behavior and its ability to accurately identify DNS tunneling activities.

\par Briefly, the decision boundary analysis further corroborates DNS Sentinel's superior performance in DNS tunneling detection, highlighting its capability to effectively discern between legitimate and malicious domain activities based on intricate features such as entropy and domain length.

\section{Conclusion and Future Work}
\par To summarize, the nature of DNS tunneling attacks presents a significant challenge to network security, exploiting DNS protocols to bypass traditional security measures and exfiltrate sensitive data covertly. As cyber threats evolve, the need for advanced detection mechanisms becomes paramount to safeguarding network infrastructures against sophisticated malicious activities.

\par Various research efforts have explored different applications.Approaches to DNS tunneling detection, machine exploitation learning algorithms, deep learning models and statistical
methods. Although some studies focused on specific classes
Screens like Naive Bayes and Support Vector Machines and others
We delved into ensemble techniques and neural networks.
Improve detection accuracy. Especially existing solutions
showed varying levels of success, but many fell short
especially in achieving optimum performance in balancing
detection rates and false positives

\par In response to the challenges posed by DNS tunneling attacks, we introduced DNS Sentinel, a sophisticated detection framework designed to augment DNS security through intelligent tunnel detection. Leveraging a multifaceted approach, DNS Sentinel incorporates a diverse array of machine learning classifiers tailored to handle the nuances of DNS traffic. The system operates within existing network frameworks, providing real-time analysis and response capabilities essential for mitigating DNS-based security risks.

\par DNS Sentinel's architecture encompasses several stages of processing, including data preparation, model training, evaluation, and deployment. The system employs feature-rich datasets for both training and testing, ensuring a balanced representation of benign and malicious domain names. By utilizing advanced machine learning algorithms and neural networks, DNS Sentinel analyzes DNS query attributes such as query length, frequency, entropy, and lexical features to identify patterns indicative of tunneling activities.

\par The evaluation of DNS Sentinel demonstrated its superior performance with an accuracy and a perfect recall rate for tunneling detection, capturing all malicious activities without sacrificing accuracy. In comparison, traditional classifiers like Multinomial and Bernoulli Naive Bayes showed limited effectiveness, while more complex models, such as Multi-Layer Perceptron and Linear Support Vector Machine, displayed less impressive results. Additionally, the decision boundary analysis validated DNS Sentinel's effectiveness, underscoring its capability to distinguish between benign and malicious DNS traffic based on detailed features.

\par In conclusion, DNS Sentinel offers a comprehensive and efficient approach to enhancing DNS security. Through its innovative design, incorporating advanced machine learning techniques and tailored feature analysis, DNS Sentinel achieves admirable accuracy in identifying malicious tunneling activities while minimizing false positives. As cyber threats continue to evolve, DNS Sentinel's adaptive framework positions it as a critical tool in fortifying network defenses against DNS-based attacks, ensuring the integrity and security of network infrastructures in an increasingly interconnected digital landscape.

\subsection{Future Work}
Future work will focus on refining the model to handle more complex and evolving DNS tunneling techniques. Incorporating real-time data analysis and continuous learning mechanisms could further enhance the system's detection capabilities and adaptability to new threats. Additionally, expanding the dataset with a wider variety of benign and malicious domain names can improve the model's generalization and accuracy. Collaboration with cybersecurity experts and network administrators to integrate DNS Sentinel into broader network security solutions will also be explored to provide a comprehensive defense against DNS-based threats.

\end{document}